\documentclass{article}

\usepackage{amsmath}
\usepackage{multicol}
\usepackage{caption}
\usepackage{amssymb}
\usepackage[preprint]{neurips_2025}
\usepackage{algorithm}
\usepackage[noend]{algpseudocode}
\usepackage{multirow}
\usepackage{caption}
\usepackage[utf8]{inputenc} 
\usepackage[T1]{fontenc}    
\usepackage{hyperref}       
\usepackage{url}            
\usepackage{booktabs}       
\usepackage{amsfonts}       
\usepackage{nicefrac}       
\usepackage{microtype}      
\usepackage{xcolor}         
\usepackage{graphicx}

\title{Accelerating Protein Molecular \\ Dynamics Simulation with DeepJump}

\author
{Allan dos Santos Costa\\
MIT Center for Bits and Atoms\\
\And
Manvitha Ponnapati\\
MIT Center for Bits and Atoms\\
\And
Dana Rubin\\
MIT Center for Bits and Atoms\\
\And
Tess Smidt\\
MIT Atomic Architects\\
\And
Joseph Jacobson\\
MIT Center for Bits and Atoms\\
}
\begin{document}

\maketitle

\begin{abstract}
Unraveling the dynamical motions of biomolecules is essential for bridging their structure and function, yet it remains a major computational challenge. Molecular dynamics (MD) simulation provides a detailed depiction of biomolecular motion, but its high-resolution temporal evolution comes at significant computational cost, limiting its applicability to timescales of biological relevance. Deep learning approaches have emerged as promising solutions to overcome these computational limitations by learning to predict long-timescale dynamics. However, generalizable kinetics models for proteins remain largely unexplored, and the fundamental limits of achievable acceleration while preserving dynamical accuracy are poorly understood. In this work, we fill this gap with DeepJump, an Euclidean-Equivariant Flow Matching-based model for predicting protein conformational dynamics across multiple temporal scales. We train DeepJump on trajectories of the diverse proteins of mdCATH, systematically studying our model's performance in generalizing to long-term dynamics of fast-folding proteins and characterizing the trade-off between computational acceleration and prediction accuracy. We demonstrate the application of DeepJump to ab initio folding, showcasing prediction of folding pathways and native states. Our results demonstrate that DeepJump achieves significant $\approx$1000$\times$ computational acceleration while effectively recovering long-timescale dynamics, providing a stepping stone for enabling routine simulation of proteins.
\end{abstract}

\section{Introduction}

Proteins form the functional infrastructure of biological systems, performing complex actions through intricate dynamic reconfiguration \citep{nelson2008lehninger}. Uncovering protein motion is hence essential to elucidating the mechanisms of biological processes, and a key step towards developing strategies to combat disease at the molecular level. Classical Molecular Dynamics (MD) simulation describes the kinetics of molecules by integrating the Newtonian equations of motion yielded from atomic force-fields \citep{schlick2010molecular}. However, high-frequency motion components often limit the practical timestep of first-principles simulation, making relevant biological timescales computationally prohibitive to achieve \citep{freddolino2010challenges}.

In contrast, Deep Learning models have demonstrated remarkable success in resolving challenging prediction tasks of protein thermodynamics, such as in structure prediction \citep{jumper2021highly} and in ensemble distribution generation \citep{jing2024alphafold, lewis2025scalable}. Still, while existing large-scale models enable generalizable prediction of static basin or equilibrium states, generalizable kinetics remains a challenging frontier for modeling biological processes, as it requires training and evaluation across vast conformational phase spaces through computationally expensive simulations.

In this work, we hypothesize that training on short, structurally diverse trajectories can successfully capture generalizable dynamical behavior. To test this hypothesis, we develop DeepJump, a generative model that combines flow matching with equivariant neural networks to model protein conformational transitions. We train our model on the structurally diverse mdCATH \citep{mirarchi2024mdcath} dataset, evaluating it on extended microsecond simulations of fast-folding proteins \citep{lindorff2011fast, majewski2023machine}. We show how the learned simulator successfully generalizes beyond its training timescales to reproduce long-term protein dynamics while achieving orders-of-magnitude computational speedup. We analyze the trade-offs between acceleration and simulation quality across different model capacities and temporal jump sizes. Finally, we demonstrate the utility of our model in performing ab initio folding simulations from extended conformations to native states.

\begin{figure}[t]
    \centering
    \includegraphics[width=0.95\linewidth]{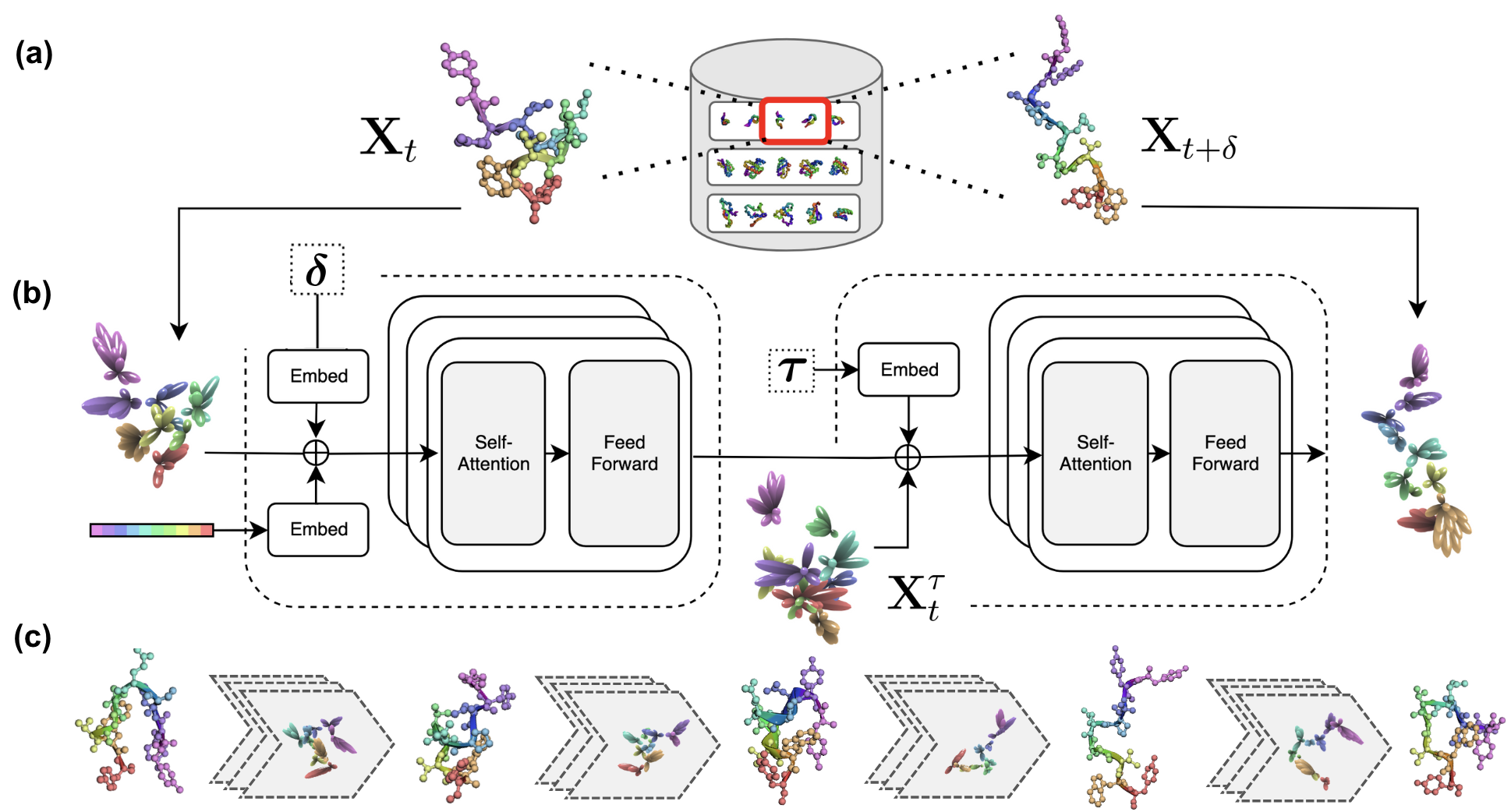}
    \caption{\textbf{DeepJump Architecture and Training}. (a) Training data consists of diverse protein trajectory snapshots from the mdCATH dataset, providing structural diversity across different protein folds. (b) Our model uses a two-stage architecture with a current-step conditioner and a generative transport network to predict the next 3D state. (c) Our generative sampler produces long trajectories by iteratively applying learned large-scale conformational moves to efficiently sample protein conformational space.}
    \label{fig:main}
\end{figure}

\section{Related Work}

Diverse deep learning efforts aim to reproduce MD and tackle its fundamental bottlenecks. Machine learning force fields (MLFFs) like ANI \citep{smith2017ani} NequIP \citep{batzner20223} and MACE \citep{batatia2022mace} have demonstrated remarkable accuracy in reproducing potentials while maintaining computational efficiency. However, MLFFs are constrained by the timestep limitations, as they still require integration of the full atomic equations of motion around femtosecond resolution. Recent breakthroughs have instead turned to generative models, with approaches like AlphaFlow \citep{jing2024alphafold} and BioEmu \citep{lewis2025scalable} training over trajectory data to generate Boltzmann ensembles. Still, ensemble models fail to capture the whole picture of dynamics, as trajectories are needed for mechanistic understanding. In further development towards capturing kinetics,
recent models \citep{ito,li2024f3low,jing24generative} enable sampling of dynamics trajectories with large steps, suggesting a pathway for accelerated MD simulation where the large time that is skipped outweighs the cost of evaluating the neural network. In this work, we build upon EquiJump \citep{costa2024equijump} and JAMUN \citep{daigavane2025jamunbridgingsmoothedmolecular}, leveraging equivariant neural networks \citep{geiger2022e3nn} for generating proximal conformational states. We investigate multiple step sizes, as in ITO \citep{ito}, while utilizing the sampling efficiency of Guided Flow Matching \citep{lipman2022flow}, as in F$^3
$low \citep{li2024f3low}. We extend this line of work to investigate acceleration-accuracy trade-offs when reproducing the dynamics of fast-folding proteins, demonstrating generalization of dynamics across both different proteins and conformational phase space, and benchmarking the limits of practical applicability for large-scale protein simulations.
\section{Methods}

\begin{figure}[t]
    \centering    \includegraphics[width=1.0\linewidth]{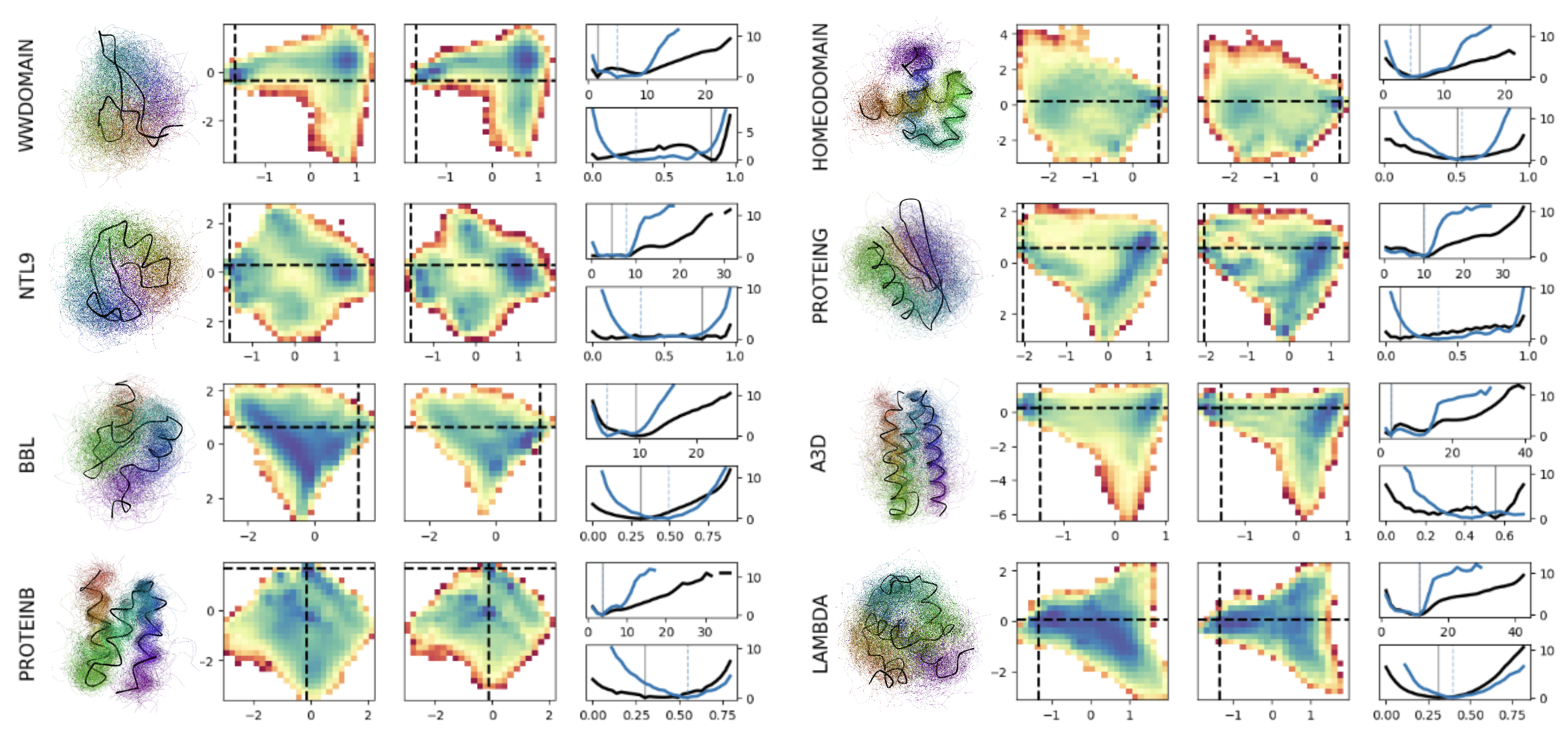}
    \caption{\textbf{Dynamics Generalization to the Fast Folder Proteins}. Ensemble visualization and free energy landscapes in TIC space (Appendix \ref{appx:equilibration}) comparing DeepJump simulations with reference molecular dynamics. Crystal 3D structure is shown in black. Free energy profiles for RMSD (top) and especially fraction of native contacts (FNC) (bottom) demonstrate strong agreement between model (blue) and reference data (black). The model successfully captures the main conformational basins and transition pathways of the fast-folding proteins.}
    \label{fig:fe}
\end{figure}

\subsection{Generative Model}

We follow EquiJump \citep{costa2024equijump} and model a protein $(\mathbf R, \mathbf X)$ as a length $N$ sequence $\mathbf R \in \mathcal R^N$ and a 3D gas of geometric features $\mathbf X = (\mathbf P, \mathbf V)$, where $\mathbf P \in \mathbb  R^{N\times3}$ are coordinates of $\mathbf C_\alpha$ atoms and $\mathbf V \in \mathbb R^{N\times13\times3}$ are 3D features listing for each residue the relative position of heavy atoms in relation to the $\mathbf C_\alpha$. We train the model using trajectory data $[\mathbf X_t ]^L_{t=1}$. Given a starting point $t$ and a jump time $\delta$, we take state transition $(\mathbf X_t, \mathbf X_{t+\delta})$ as the endpoints of a Conditional Flow Matching/Stochastic Interpolant \citep{li2024f3low, lipman2022flow, albergo2023stochastic} model that maps from a noised source state $\rho_0=\rho(\mathbf X_t + \mathbf Z)$, where $\mathbf Z \sim \mathcal N$, into a future time step $\rho_1=\rho(\mathbf X_{t+\delta}|\mathbf X_t)$ (Figure \ref{fig:main}.b). We learn a model to predict the final state directly, $\mathbf{\hat{X}^1}(\mathbf X^\tau|\tau) \approx \mathbf X^1$, and in sampling reparameterize via $
b(\mathbf X^\tau|\tau) =  \frac{1}{(1-\tau)} \big (\hat{\mathbf X}^1(\mathbf X^\tau|\tau)- \mathbf X_t \big )$ \citep{jing2024alphafold}.


\subsection{Architecture and Training}

Our architecture consists of two main stages (Figure \ref{fig:main}.b). First, a conditioning encoder computes $\mathbf H_t =\mathbf f_{\textrm{cond}}(\mathbf X_t, \mathbf R, \delta)$ from the current structural state $\mathbf X_t$, sequence $\mathbf R$, and jump time $\delta$. Second, a transport network $\mathbf f_{\textrm{transp}}(\mathbf X^\tau|\tau,\mathbf H_t)$ iteratively updates the latent state $\mathbf X^\tau$ to generate a new configuration. Both networks use Euclidean-equivariant architectures \citep{geiger2022e3nn} inspired by Transformer mechanisms \citep{vaswani2017attention} adapted to equivariant space (see Appendix \ref{appx:arch} for details). During training, we optimize pairwise 3D vectors between all atoms within $d=25$Å using the Huber Loss \citep{costa2024ophiuchus, huber1992robust}.

\subsection{Datasets}
To ensure the generalization power of our model, we train it using the diverse structures of the mdCATH dataset \citep{mirarchi2024mdcath}. This dataset consists of all-atom systems for 5,398 domains, modeled with a state-of-the-art classical force field, and simulated in five replicates of 500 ns from the crystal state, each at five temperatures from 320 K to 450 K. While this dataset encompasses a broad range of different proteins, it is not sufficient for capturing long-timescale dynamical behavior and equilibrium properties due to its limited simulation time per trajectory. Instead, for evaluating our dynamics we test our model on the dataset of 12 fast-folder proteins of \citep{majewski2023machine} based on \citep{lindorff2011fast}. In contrast to the training data, this set provides significant accumulated simulation time with thousands of trajectory snippets across the phase space, enabling precise estimation of dynamical variables and asymptotic behavior.

\section{Results}

\begin{table}[t]
\caption{\textbf{Model Performance across Jump Sizes}. We fit Markov State Models (MSM) to the transitions between TIC-based clusters, and compare obtained MSMs from reference and from learned models. Results are averaged over the fast-folding proteins. We use Jensen Shannon Divergence to measure distribution distance for stationary distributions and transition matrix (averaging over rows), and absolute differences otherwise. To estimate folding metrics, we compare energetics and timescales between  clusters corresponding to the $\alpha$-helix state and the crystal state.}
\centering

\resizebox{\textwidth}{!}{%
\begin{tabular}{@{}rrrrrrrrrr@{}}
\toprule
\textbf{$\delta$ (ns)}                    & \multicolumn{3}{c}{\textbf{1}}           & \multicolumn{3}{c}{\textbf{10}}          & \multicolumn{3}{c}{\textbf{100}}         \\ \cmidrule(l){2-4} \cmidrule(l){5-7} \cmidrule(l){8-10} 
\textbf{Model Dimensionality}             & \textbf{32} & \textbf{64} & \textbf{128} & \textbf{32} & \textbf{64} & \textbf{128} & \textbf{32} & \textbf{64} & \textbf{128} \\ \midrule
\textbf{Stationary Distribution Distance (bits)} & 0.18        & 0.06        & 0.07         & 0.27        & 0.11        & 0.17         & 0.29        & 0.24        & 0.31         \\
\textbf{Folding $\Delta$G Error ($k_b T$)}          & 3.02        & 1.24        & 1.14         & 3.64        & 2.05        & 1.88         & 5.11        & 3.37        & 3.24         \\
\textbf{Transition Matrix Distance (bits)}       & 0.25        & 0.28        & 0.27         & 0.45        & 0.48        & 0.46         & 0.42        & 0.46        & 0.44         \\
\textbf{Folding MFPT Error (us)}               & 69.2        & 0.3         & 0.4          & 10.5       & 6.7        & 7.0         & 190.5      & 225.0       & 37.4        \\ \bottomrule
\end{tabular}%
}
\label{tab:abl}
\end{table}

\begin{figure}[t]
    \centering
    \includegraphics[width=1.0\linewidth]{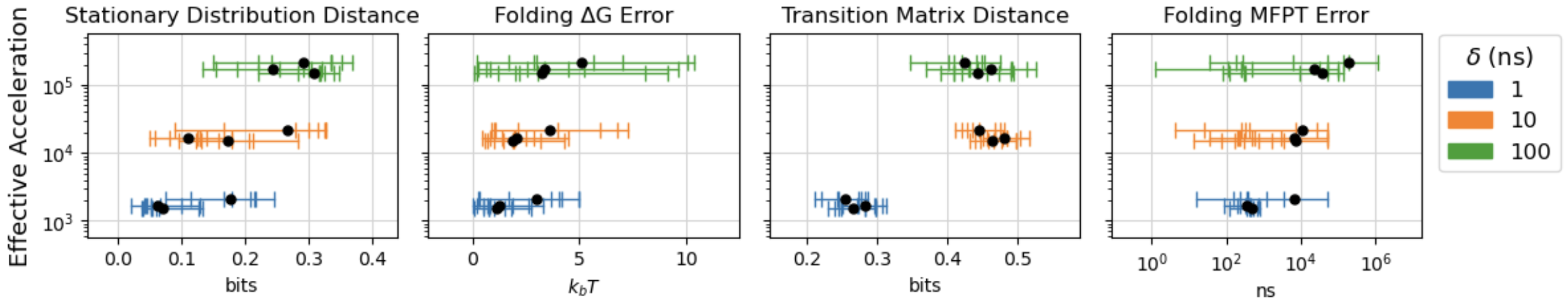}
    \caption{\textbf{Mapping Acceleration Fronts}. We investigate the tradeoff between simulation fidelity and computational speedup by varying model scale and conditioned jump size $\delta$. We find that larger jumps degrade simulation quality, with increased model capacity reducing error but only partially mitigating the effect.}
    \label{fig:accel}
\end{figure}

\subsection{DeepJump Generalizes to Fast-Folder Phase Space}

To assess DeepJump's ability to capture long-term dynamics beyond its short training trajectories, we extensively sample our model across the conformational phase space of the fast-folding proteins. For that, we employ Time-lagged Independent Component Analysis (TICA) \citep{molgedey1994separation, perez2013identification} and find clusters that represent macrostates in the reduced dimensional space. Refer to Appendix \ref{appx:equilibration} for further details. For each fast-folder, we start 1200 replicas uniformily across the clusters, and perform 1000 simulation steps. We fit a Markov State Model (MSM) to transition counts between clusters, and correct our measured observables to the MSM stationary distribution to estimate free energies (Figure \ref{fig:fe}). Analysis of the TIC free energy profiles shows that the learned simulator is able to generalize to unseen proteins and across the phase space. Similarly, while RMSD and FNC energy (Figure \ref{fig:fe}) analysis suggests a bias towards compact conformations, the model overall shows strong agreement with the reference data.

\subsection{Mapping the Frontiers of MD Acceleration}
To better understand the trade-offs between simulation accuracy and computational speedup, we analyze the MSMs constructed from simulations with different model capacities and jump step sizes, comparing them to MSMs built from the reference data. Table \ref{tab:abl} shows the quantitative comparison of MSM properties across different configurations. We present these results in condensed form in Figure \ref{fig:accel}, where we estimate effective acceleration relative to Amber force-field \citep{wang2004development} simulations (32 real s / simulation ns on A6000 \citep{amber-benchmarks}) for the Lambda protein. Our plots show that while jump size significantly impacts simulation quality, model scaling can modestly compensate for this degradation. Nevertheless, our results reveal that substantial acceleration remains achievable within acceptable quality bounds.

\subsection{Accelerating ab initio Folding}
\begin{figure}[H]
    \centering    \includegraphics[width=1.0\linewidth]{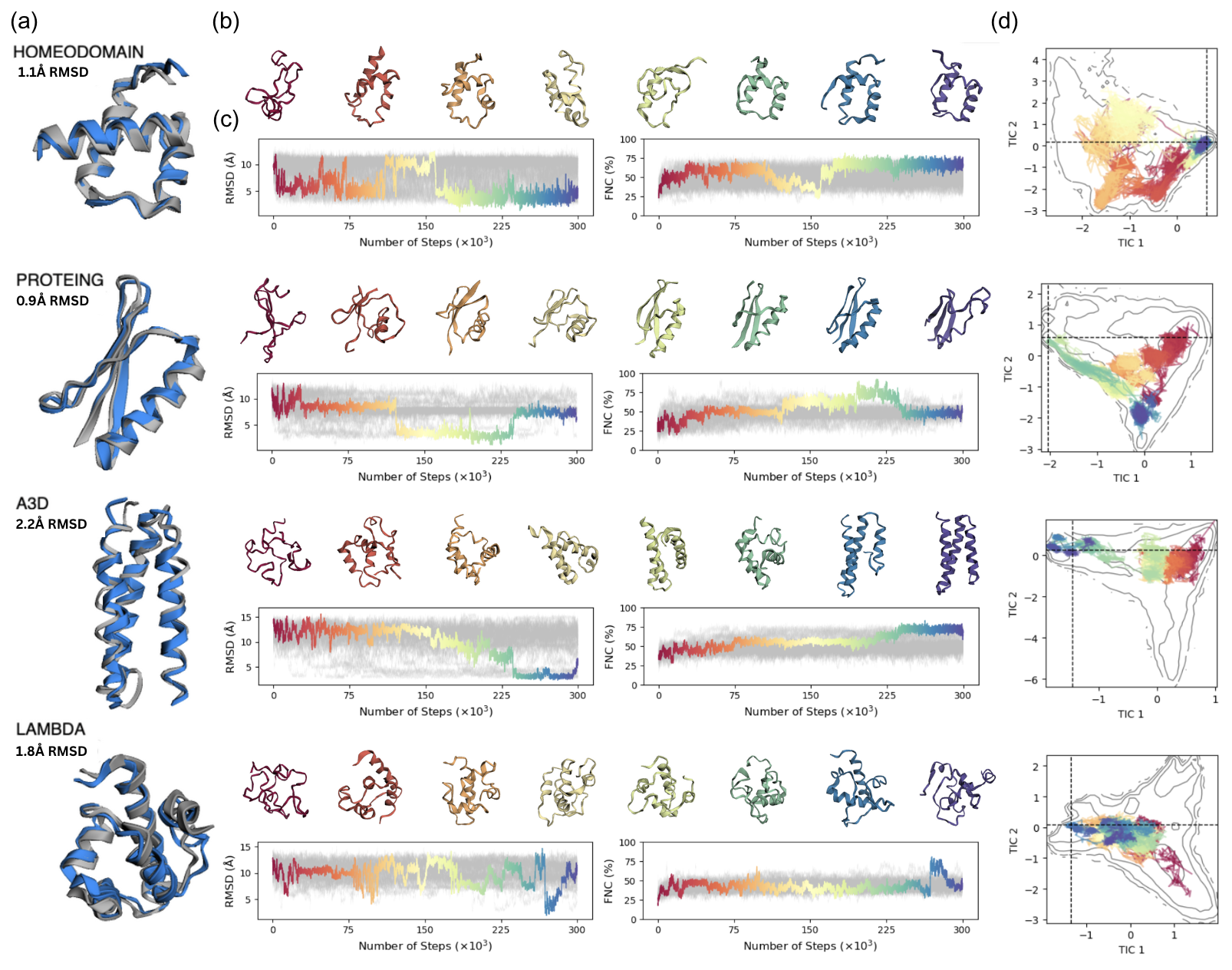}
    \caption{\textbf{Generative Simulation of Protein Folding}. We run 300k simulation steps for 64 replicas of fast-folding proteins. (a) We show best aligning sample (blue) to crystal structure (gray). (b) We highlight a trajectory achieving highest fraction of native contacts (FNC) and plot its 3D steps aligned to crystal. (c) We show the time evolution of RMSD and FNC for the replicas (gray) and highlighted trajectory (colored). (d) We plot TIC coordinates against the reference phase space contour (gray). }
    \label{fig:abinitio}
\end{figure}

To evaluate DeepJump in a practical application, we investigate its performance on the challenging task of ab initio protein folding. For each fast-folder, we start 64 replicas from an extended $\beta$-sheet state and perform 300k simulation steps. In Figure \ref{fig:abinitio}, we show folding trajectories and sampled structures with the closest match to the native state. Investigation of the curves reveals that our simulation successfully captures smooth folding pathways with physically realistic conformational transitions. Table \ref{tab:abinitio} compares the performance of models using different jump sizes $\delta$. We find that folding success varies with step size: models with 1ns and 10ns steps achieving the highest quality results, while 100ns steps fail to fold some proteins entirely. This is due to the increasing difficulty of accurately modeling large conformational transitions over extended time intervals, where smaller steps enable the model to capture rare barrier-crossing events and intermediate states that are crucial for successful folding, while larger jumps may bypass conformational pathways or become trapped in local minima. Refer to Appendix \ref{appx:abinitio} for further discussion.

\begin{table}[t]
\caption{\textbf{Folding from Scratch}. We quantify folding success by identifying trajectories that reach the TICA-based cluster corresponding to the native crystal state. For the proteins that fold, we count success per replica and Mean First Passage Time (MFPT) from beta sheet to folded state. Results are averaged over the fast-folders.}
\centering
\resizebox{0.62\textwidth}{!}{%
\begin{tabular}{@{}rrrr@{}}
\toprule
\textbf{$\delta$ (ns)}   & \textbf{1}      & \textbf{10}     & \textbf{100}      \\ \midrule
Proteins Folded (\%) $\uparrow$    & \textbf{100.00} & \textbf{100.00} & 62.50             \\
Replicas Folded (\%) $\uparrow$    & 50.59           & \textbf{61.13}  & 57.23             \\
Mininum Crystal RMSD (Å) $\downarrow$ & \textbf{1.54}   & 1.64            & 2.35              \\
Maximum FNC (\%)  $\uparrow$    & 86.40           & \textbf{87.10}  & 77.96             \\
MFPT ($\times 10^3$ Model Steps)  $\downarrow$     & 97.3        & 81.8        & 12.7 \\ \bottomrule
\end{tabular}%
}
\label{tab:abinitio}
\end{table}

\subsection{Model Limitations}
While our model successfully generalizes to most fast-folder proteins, we found limits to its applicability across all systems. In particular, we found that it fails on proteins much smaller than those in the training data (e.g., Chignolin or Trp-Cage), generating chemically invalid states. We also highlight bias (Figure \ref{fig:fe}) toward globular conformations and basin states, as the training data predominantly consists of well-folded protein domains, limiting the model's ability to capture disordered or extended conformational states that are often crucial to pathway modeling. Finally, our modeling assumes standard residues, which prevents application to proteins with non-standard amino acids (e.g., fast-folder Villin).
\section*{Conclusion}

We have presented DeepJump, a generative model that leverages flow matching and equivariant neural networks to accelerate protein molecular dynamics simulations by learning conformational transitions from diverse trajectories across timescales. Our approach successfully reproduces key dynamical properties of fast-folding proteins, including realistic folding pathways and equilibrium distributions, while achieving orders-of-magnitude acceleration compared to traditional force-field simulations. Through analysis of acceleration fronts, we demonstrate important trade-offs between simulation speed and accuracy, where larger jump sizes provide greater computational speedup at the cost of simulation quality, with model scaling offering compensation. Additionally, in ab initio folding experiments, we show that the model can successfully fold proteins from extended conformations to native-like states, with folding success depending on the chosen temporal step size. In conclusion, DeepJump represents a promising step toward practical machine learning-accelerated molecular simulations, offering a path to building simulators to previously inaccessible timescales.

\section*{Acknowledgments}
We thank Ilan Mitnikov, Emine Kucukbenli, Franco Pellegrini, Mario Geiger and Mit Kotak for insightful discussions. This research was made possible through the support of DARPA (AI‑BTO SCA‑24‑01), the Eleven Eleven Foundation, the Center for Bits and Atoms, and the MIT Media Lab Consortium.


\bibliographystyle{abbrvnat}
\bibliography{ref.bib}

\begin{thebibliography}{30}
\providecommand{\natexlab}[1]{#1}
\providecommand{\url}[1]{\texttt{#1}}
\expandafter\ifx\csname urlstyle\endcsname\relax
  \providecommand{\doi}[1]{doi: #1}\else
  \providecommand{\doi}{doi: \begingroup \urlstyle{rm}\Url}\fi

\bibitem[Albergo et~al.(2023)Albergo, Boffi, and Vanden-Eijnden]{albergo2023stochastic}
M.~S. Albergo, N.~M. Boffi, and E.~Vanden-Eijnden.
\newblock Stochastic interpolants: A unifying framework for flows and diffusions, 2023.

\bibitem[Batatia et~al.(2022)Batatia, Kovacs, Simm, Ortner, and Cs{\'a}nyi]{batatia2022mace}
I.~Batatia, D.~P. Kovacs, G.~Simm, C.~Ortner, and G.~Cs{\'a}nyi.
\newblock Mace: Higher order equivariant message passing neural networks for fast and accurate force fields.
\newblock \emph{Advances in Neural Information Processing Systems}, 35:\penalty0 11423--11436, 2022.

\bibitem[Batzner et~al.(2022)Batzner, Musaelian, Sun, Geiger, Mailoa, Kornbluth, Molinari, Smidt, and Kozinsky]{batzner20223}
S.~Batzner, A.~Musaelian, L.~Sun, M.~Geiger, J.~P. Mailoa, M.~Kornbluth, N.~Molinari, T.~E. Smidt, and B.~Kozinsky.
\newblock E (3)-equivariant graph neural networks for data-efficient and accurate interatomic potentials.
\newblock \emph{Nature communications}, 13\penalty0 (1):\penalty0 2453, 2022.

\bibitem[Costa et~al.(2024{\natexlab{a}})Costa, Mitnikov, Geiger, Ponnapati, Smidt, and Jacobson]{costa2024ophiuchus}
A.~D.~S. Costa, I.~Mitnikov, M.~Geiger, M.~Ponnapati, T.~Smidt, and J.~Jacobson.
\newblock Ophiuchus: Scalable modeling of protein structures through hierarchical coarse-graining {SO}(3)-equivariant autoencoders.
\newblock In \emph{ICLR 2024 Workshop on Generative and Experimental Perspectives for Biomolecular Design}, 2024{\natexlab{a}}.
\newblock URL \url{https://openreview.net/forum?id=hnhhCfYeWU}.

\bibitem[Costa et~al.(2024{\natexlab{b}})Costa, Mitnikov, Pellegrini, Daigavane, Geiger, Cao, Kreis, Smidt, Kucukbenli, and Jacobson]{costa2024equijump}
A.~D.~S. Costa, I.~Mitnikov, F.~Pellegrini, A.~Daigavane, M.~Geiger, Z.~Cao, K.~Kreis, T.~Smidt, E.~Kucukbenli, and J.~Jacobson.
\newblock Equijump: Protein dynamics simulation via so (3)-equivariant stochastic interpolants.
\newblock \emph{arXiv preprint arXiv:2410.09667}, 2024{\natexlab{b}}.

\bibitem[Daigavane et~al.(2025)Daigavane, Vani, Davidson, Saremi, Rackers, and Kleinhenz]{daigavane2025jamunbridgingsmoothedmolecular}
A.~Daigavane, B.~P. Vani, D.~Davidson, S.~Saremi, J.~Rackers, and J.~Kleinhenz.
\newblock Jamun: Bridging smoothed molecular dynamics and score-based learning for conformational ensembles, 2025.
\newblock URL \url{https://arxiv.org/abs/2410.14621}.

\bibitem[Exxact Corp.()]{amber-benchmarks}
Exxact Corp.
\newblock Amber 24 nvidia gpu benchmarks.
\newblock \url{https://www.exxactcorp.com/blog/molecular-dynamics/amber-molecular-dynamics-nvidia-gpu-benchmarks}, 2024.

\bibitem[Freddolino et~al.(2010)Freddolino, Harrison, Liu, and Schulten]{freddolino2010challenges}
P.~Freddolino, C.~Harrison, Y.~Liu, and K.~Schulten.
\newblock Challenges in protein 524 folding simulations: timescale, representation, and analysis.
\newblock \emph{Nat. Phys}, 6\penalty0 (751-758):\penalty0 525, 2010.

\bibitem[Geiger and Smidt(2022)]{geiger2022e3nn}
M.~Geiger and T.~Smidt.
\newblock e3nn: Euclidean neural networks.
\newblock \emph{arXiv preprint arXiv:2207.09453}, 2022.

\bibitem[Huber(1992)]{huber1992robust}
P.~J. Huber.
\newblock Robust estimation of a location parameter.
\newblock In \emph{Breakthroughs in statistics: Methodology and distribution}, pages 492--518. Springer, 1992.

\bibitem[Jing et~al.(2020)Jing, Eismann, Suriana, Townshend, and Dror]{jing2020learning}
B.~Jing, S.~Eismann, P.~Suriana, R.~J. Townshend, and R.~Dror.
\newblock Learning from protein structure with geometric vector perceptrons.
\newblock \emph{arXiv preprint arXiv:2009.01411}, 2020.

\bibitem[Jing et~al.(2024{\natexlab{a}})Jing, Berger, and Jaakkola]{jing2024alphafold}
B.~Jing, B.~Berger, and T.~Jaakkola.
\newblock Alphafold meets flow matching for generating protein ensembles.
\newblock \emph{arXiv preprint arXiv:2402.04845}, 2024{\natexlab{a}}.

\bibitem[Jing et~al.(2024{\natexlab{b}})Jing, Stark, Jaakkola, and Berger]{jing24generative}
B.~Jing, H.~Stark, T.~Jaakkola, and B.~Berger.
\newblock Generative modeling of molecular dynamics trajectories.
\newblock In \emph{ICML'24 Workshop ML for Life and Material Science: From Theory to Industry Applications}, 2024{\natexlab{b}}.

\bibitem[Jumper et~al.(2021)Jumper, Evans, Pritzel, Green, Figurnov, Ronneberger, Tunyasuvunakool, Bates, {\v{Z}}{\'\i}dek, Potapenko, et~al.]{jumper2021highly}
J.~Jumper, R.~Evans, A.~Pritzel, T.~Green, M.~Figurnov, O.~Ronneberger, K.~Tunyasuvunakool, R.~Bates, A.~{\v{Z}}{\'\i}dek, A.~Potapenko, et~al.
\newblock Highly accurate protein structure prediction with alphafold.
\newblock \emph{Nature}, 596\penalty0 (7873):\penalty0 583--589, 2021.

\bibitem[Lewis et~al.(2025)Lewis, Hempel, Jim{\'e}nez-Luna, Gastegger, Xie, Foong, Satorras, Abdin, Veeling, Zaporozhets, et~al.]{lewis2025scalable}
S.~Lewis, T.~Hempel, J.~Jim{\'e}nez-Luna, M.~Gastegger, Y.~Xie, A.~Y. Foong, V.~G. Satorras, O.~Abdin, B.~S. Veeling, I.~Zaporozhets, et~al.
\newblock Scalable emulation of protein equilibrium ensembles with generative deep learning.
\newblock \emph{Science}, page eadv9817, 2025.

\bibitem[Li et~al.(2024)Li, Wang, Li, Zhang, Shao, Zheng, and Tang]{li2024f3low}
S.~Li, Y.~Wang, M.~Li, J.~Zhang, B.~Shao, N.~Zheng, and J.~Tang.
\newblock F$^3$low: Frame-to-frame coarse-grained molecular dynamics with se(3) guided flow matching, 2024.

\bibitem[Liao and Smidt(2022)]{liao2022equiformer}
Y.-L. Liao and T.~Smidt.
\newblock Equiformer: Equivariant graph attention transformer for 3d atomistic graphs.
\newblock \emph{arXiv preprint arXiv:2206.11990}, 2022.

\bibitem[Lindorff-Larsen et~al.(2011)Lindorff-Larsen, Piana, Dror, and Shaw]{lindorff2011fast}
K.~Lindorff-Larsen, S.~Piana, R.~O. Dror, and D.~E. Shaw.
\newblock How fast-folding proteins fold.
\newblock \emph{Science}, 334\penalty0 (6055):\penalty0 517--520, 2011.

\bibitem[Lipman et~al.(2022)Lipman, Chen, Ben-Hamu, Nickel, and Le]{lipman2022flow}
Y.~Lipman, R.~T. Chen, H.~Ben-Hamu, M.~Nickel, and M.~Le.
\newblock Flow matching for generative modeling.
\newblock \emph{arXiv preprint arXiv:2210.02747}, 2022.

\bibitem[Lloyd(1982)]{lloyd1982least}
S.~Lloyd.
\newblock Least squares quantization in pcm.
\newblock \emph{IEEE transactions on information theory}, 28\penalty0 (2):\penalty0 129--137, 1982.

\bibitem[Majewski et~al.(2023)Majewski, P{\'e}rez, Th{\"o}lke, Doerr, Charron, Giorgino, Husic, Clementi, No{\'e}, and De~Fabritiis]{majewski2023machine}
M.~Majewski, A.~P{\'e}rez, P.~Th{\"o}lke, S.~Doerr, N.~E. Charron, T.~Giorgino, B.~E. Husic, C.~Clementi, F.~No{\'e}, and G.~De~Fabritiis.
\newblock Machine learning coarse-grained potentials of protein thermodynamics.
\newblock \emph{Nature Communications}, 14\penalty0 (1):\penalty0 5739, 2023.

\bibitem[Mirarchi et~al.(2024)Mirarchi, Giorgino, and De~Fabritiis]{mirarchi2024mdcath}
A.~Mirarchi, T.~Giorgino, and G.~De~Fabritiis.
\newblock mdcath: A large-scale md dataset for data-driven computational biophysics.
\newblock \emph{Scientific Data}, 11\penalty0 (1):\penalty0 1299, 2024.

\bibitem[Molgedey and Schuster(1994)]{molgedey1994separation}
L.~Molgedey and H.~G. Schuster.
\newblock Separation of a mixture of independent signals using time delayed correlations.
\newblock \emph{Physical review letters}, 72\penalty0 (23):\penalty0 3634, 1994.

\bibitem[Nelson et~al.(2008)Nelson, Lehninger, and Cox]{nelson2008lehninger}
D.~L. Nelson, A.~L. Lehninger, and M.~M. Cox.
\newblock \emph{Lehninger principles of biochemistry}.
\newblock Macmillan, 2008.

\bibitem[P{\'e}rez-Hern{\'a}ndez et~al.(2013)P{\'e}rez-Hern{\'a}ndez, Paul, Giorgino, De~Fabritiis, and No{\'e}]{perez2013identification}
G.~P{\'e}rez-Hern{\'a}ndez, F.~Paul, T.~Giorgino, G.~De~Fabritiis, and F.~No{\'e}.
\newblock Identification of slow molecular order parameters for markov model construction.
\newblock \emph{The Journal of chemical physics}, 139\penalty0 (1), 2013.

\bibitem[Schlick(2010)]{schlick2010molecular}
T.~Schlick.
\newblock \emph{Molecular modeling and simulation: an interdisciplinary guide}, volume~2.
\newblock Springer, 2010.

\bibitem[Schreiner et~al.(2023)Schreiner, Winther, and Olsson]{ito}
M.~Schreiner, O.~Winther, and S.~Olsson.
\newblock Implicit transfer operator learning: Multiple time-resolution surrogates for molecular dynamics, 2023.

\bibitem[Smith et~al.(2017)Smith, Isayev, and Roitberg]{smith2017ani}
J.~S. Smith, O.~Isayev, and A.~E. Roitberg.
\newblock Ani-1: an extensible neural network potential with dft accuracy at force field computational cost.
\newblock \emph{Chemical science}, 8\penalty0 (4):\penalty0 3192--3203, 2017.

\bibitem[Vaswani et~al.(2017)Vaswani, Shazeer, Parmar, Uszkoreit, Jones, Gomez, Kaiser, and Polosukhin]{vaswani2017attention}
A.~Vaswani, N.~Shazeer, N.~Parmar, J.~Uszkoreit, L.~Jones, A.~N. Gomez, {\L}.~Kaiser, and I.~Polosukhin.
\newblock Attention is all you need.
\newblock \emph{Advances in neural information processing systems}, 30, 2017.

\bibitem[Wang et~al.(2004)Wang, Wolf, Caldwell, Kollman, and Case]{wang2004development}
J.~Wang, R.~M. Wolf, J.~W. Caldwell, P.~A. Kollman, and D.~A. Case.
\newblock Development and testing of a general amber force field.
\newblock \emph{Journal of computational chemistry}, 25\penalty0 (9):\penalty0 1157--1174, 2004.

\end{thebibliography}


\newpage
\appendix

\newpage
\section*{Appendix}

\section{Architecture and Training Details}
\label{appx:arch}

\newcommand{\lm}{{l_{\max}}}

To operate efficiently on large proteins, we adapt a form of the attention mechanism to handle equivariant vectors (Algorithm \ref{alg:att}). Drawing from GVP \citep{jing2020learning}, our feedforward layers (Algorithm \ref{alg:ff}) interact vector ($l=1$) and scalar ($l=0$) features by incorporating vector norms into scalar processing, and gating vectors through scalars. All network modules incorporate residual connections and equivariant LayerNorm \citep{liao2022equiformer} for stable training. We train all models with 6 layers for the conditional network and 6 layers for the transport network. Models of dimensionality $H$ of 32, 64 and 128 have 260k, 1M and 4M parameters respectively. We train our models on 4 A6000 machines. Models are trained for 500k steps  with batch size of 128 and crop length of 256. We use the Adam optimizer with learning rate decaying linearly from $5\times10^{-3}$ to $3\times10^{-3}$, and gradient norm clip of 0.1.

\begin{minipage}[t]{0.49\textwidth}
\begin{algorithm}[H]
\caption{DeepJump Self-Attention}\label{alg:att}
\begin{algorithmic}[1]
\Require Tensor Cloud $(\mathbf V, \mathbf P)$
\Require Number of Heads  $N_h$ 
\Require Spatial Distance Embedding  $f_d$ 
\Require Sequence Distance Embedding  $f_s$ 


\State $\mathbf k, \mathbf q, \mathbf v$ $\leftarrow \textrm{Linear}^{3  \times (H \times N_h)}(\mathbf V)$
\State $ \mathbf  v_{ijh} \leftarrow\mathbf  v_{jh} \oplus Y(\mathbf P_i - \mathbf P_j)$
\State $\mathbf  s_{ijh}$ $\leftarrow \mathbf  k_{ih} \cdot \mathbf  q_{jh}  + f_s(i-j) +  f_d(||\mathbf P_i - \mathbf P_j||_2)$
\State $\mathbf  v_{ih}$ $\leftarrow \sum_j^N  \textrm{Softmax}(\mathbf  s_{ijh}$) $\cdot\mathbf   v_{ijh}$
\State $\mathbf V'_i$ $\leftarrow$ Linear$^{H}(\bigoplus^{N_h}_{h} \mathbf  v_{ih})$

\State \Return{$(\mathbf V, \mathbf P)$}
\end{algorithmic}
\label{alg:spatial-conv}
\end{algorithm}

\end{minipage}
\hfill
\begin{minipage}[t]{0.49\textwidth}
\begin{algorithm}[H]
\caption{DeepJump FeedForward}\label{alg:ff}
\begin{algorithmic}[1]
\Require Tensor Cloud $(\mathbf P, \mathbf V)$
\Require Scaling Factor  $F$ 
\Require Activation Function $\sigma$ 

\State $\mathbf V^0_{\shortparallel}, \mathbf V^{0}_\times \leftarrow \mathrm {Linear}^{2\times (F\times H)} (\mathbf V^0)$
\State $\mathbf V^1_\shortparallel, \mathbf V^{1} _\times\leftarrow \mathrm {Linear}^{2\times (F\times H)} (\mathbf V^1)$
\State $\mathbf V_\shortparallel \leftarrow \sigma(\mathbf V^0_\shortparallel ) \oplus \sigma(\mathbf V^1_\shortparallel)$
\State $\mathbf V_\times \leftarrow \big ( \sigma(\mathbf V^1_\times)  \cdot \mathbf V^0_\times \big ) \oplus \big ( \sigma(\mathbf V^0_\times)  \cdot \mathbf V^1_\times \big )  $
\State $\mathbf V \leftarrow \textrm{Linear}^{H}(\mathbf V_\shortparallel \oplus \mathbf V_\times)$
\State \Return $(\mathbf V, \mathbf P)$
\end{algorithmic}
\label{alg:self-interaction}
\end{algorithm}
\end{minipage}

In all models, we use $N_h = 4$, $F = 2$, $\sigma=\tanh(||\cdot||^2_2)$ where the norm squared is computed multipliticy-wise. We embed sequence distance with $f_s = \textrm{Embed}(\min(\max (i-j + k_s, 2k_s), 0))$ with $k_s=32$. We use a gaussian basis function for the spatial distance embed $f_d$.

\section{Markov State Model and Dynamical Equilibration}
\label{appx:equilibration}
We fit 4 Time-lagged Independent Components (TICs)  \citep{perez2013identification} to the reference data with a lag time of 10 ns. To partition the TIC space, we apply k-means clustering \citep{lloyd1982least} with 32 clusters. We construct a Markov State Model from transition counts with lag time of 1ns between clusters and estimate its stationary distribution. We correct sampling densities by reweighting each cluster according to the ratio of its stationary probability to its observed frequency in our simulations.  When comparing the MSM transition matrices of learned models to reference data, we compare the matrix to the $\delta
$-th power to account for the different temporal resolutions.

\newpage
\section{Extended ab initio Plots}
\label{appx:abinitio}

\begin{figure}[H]
    \centering    \includegraphics[width=1.0\linewidth]{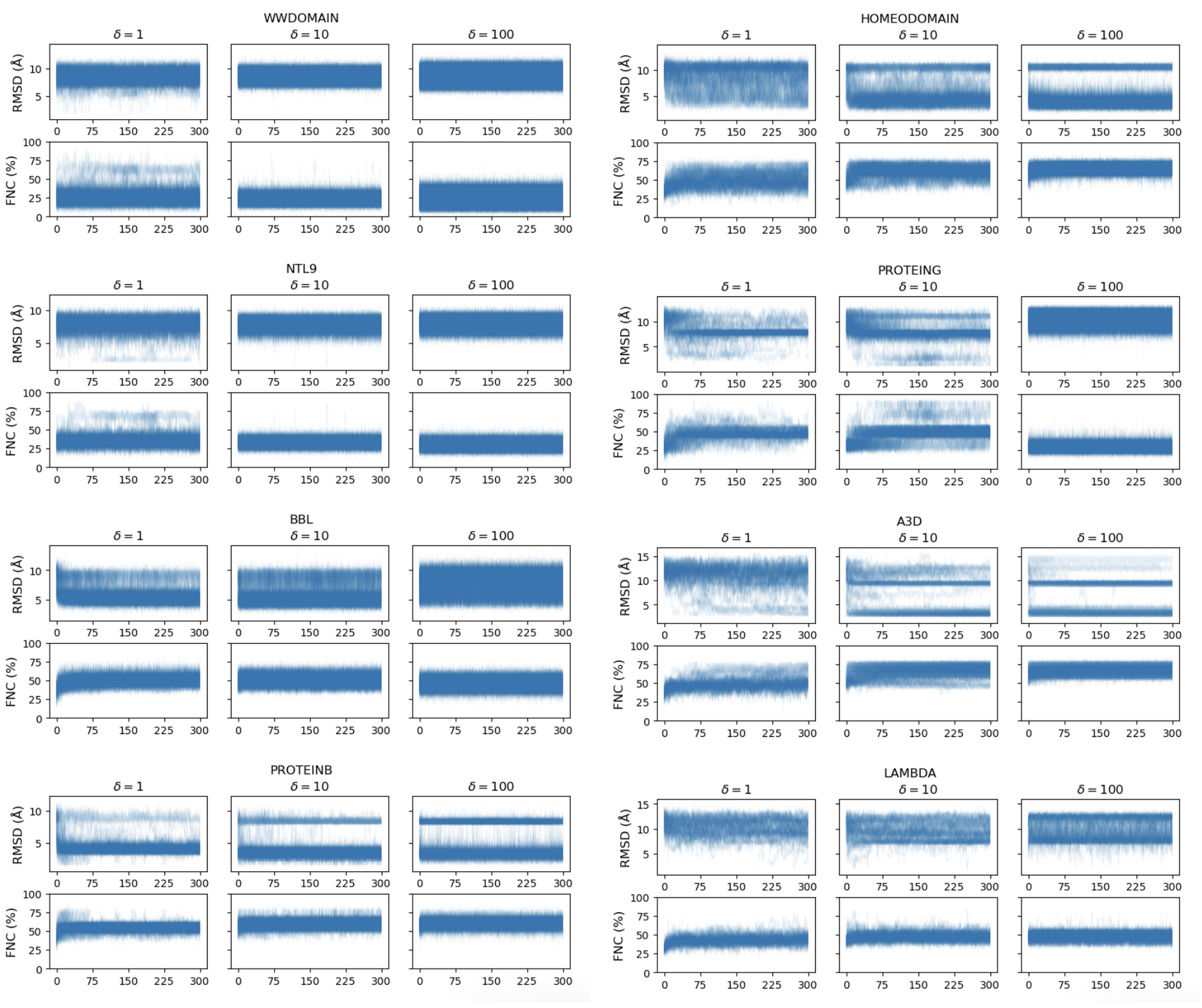}
    \caption{\textbf{Extended Plots for Folding Simulation}. }
    \label{fig:exab}
\end{figure}

In Figure \ref{fig:exab} we plot the evolution of RMSD and FNC over 300k model simulation steps for different learned steps sizes $\delta$. We observe that $\delta = 1$ ns shows the most consistent folding success across proteins, frequently reaching native basins and maintaining stability whereas $\delta = 10$ ns demonstrates intermediate stability (as seen in WW domain and NTL9). While $\delta = 100$ ns manages to fold several proteins, it fails to sample high-energy transition pathways that require fine-grained conformational sampling, such as the complex folding routes observed in NTL9 and Protein G.

\end{document}